\documentclass[letterpaper, 10 pt, conference]{ieeeconf}  
\IEEEoverridecommandlockouts                              

\usepackage{array}
\usepackage{booktabs}
\usepackage{tabularx}
\usepackage{multirow}
\usepackage{amsmath,amsfonts}
\usepackage{algorithmic}
\usepackage{algorithm}
\usepackage{mathtools}
\usepackage{amsmath}
\usepackage{indentfirst}
\usepackage{titlesec}
\usepackage{array}
\usepackage{textcomp}
\usepackage{stfloats}
\usepackage{url}
\usepackage{verbatim}
\usepackage{graphicx}
\usepackage{subcaption}
\usepackage{cite}
\usepackage{listings}
\usepackage{color}

\usepackage[utf8]{inputenc} 

\title{\LARGE \bf
Data on the Move:  
Traffic-Oriented Data Trading Platform Powered by AI Agent with Common Sense}

\author{
Yi Yu$^{1}$, Shengyue Yao$^{1,\dag}$, Tianchen Zhou$^{1,2}$, Yexuan Fu$^{1,2}$, Jingru Yu$^{1}$, \\
Ding Wang$^{1}$, Xuhong Wang$^{1}$, Cen Chen$^{2}$, and Yilun Lin$^{1,*}$, \textit{Member, IEEE}
\thanks{This work is supported by the Shanghai Artificial Intelligence Laboratory with National KeyR\&D Program of China” No. 2022ZD0160104.}
\thanks{* for the corresponding author, $\dag$ for the co-first author}
\thanks{$^{1}$Yi Yu (yuyi@pjlab.org.cn), Shengyue Yao (yaoshengyue@pjlab.org.cn), Jingru Yu(yujingru@pjlab.org.cn), Ding Wang (wangding@pjlab.org.cn) and Yilun Lin (linyilun@pjlab.org.cn) are with Urban Computing Lab, Shanghai Artificial Intelligence Laboratory, Shanghai, China.}
\thanks{$^{2}$ Tianchen Zhou (tianchen@stu.ecnu.edu.cn), Yexuan Fu (cynthiafyx@gmail.com), and Cen Chen(cenchen@dase.ecnu.edu.cn) are with the School of Data Science and Engineering, East China Normal University. 
 are with at the School of Data Science and Engineering, East China Normal University, Shanghai, China}
 \thanks{The code of the DTM platform: https://github.com/FairsLab/DTM}
}

\begin{document}

\maketitle

\thispagestyle{empty}
\pagestyle{empty}

\begin{abstract}
In the digital era, data has become a pivotal asset, advancing technologies such as autonomous driving. Despite this, data trading faces challenges like the absence of robust pricing methods and the lack of trustworthy trading mechanisms. To address these challenges, we introduce a traffic-oriented data trading platform named Data on The Move (DTM), integrating traffic simulation, data trading, and Artificial Intelligent (AI) agents. The DTM platform supports evident-based data value evaluation and AI-based trading mechanisms. Leveraging the common sense capabilities of Large Language Models (LLMs) to assess traffic state and data value, DTM can determine reasonable traffic data pricing through multi-round interaction and simulations. Moreover, DTM provides a pricing method validation by simulating traffic systems, multi-agent interactions, and the heterogeneity and irrational behaviors of individuals in the trading market. Within the DTM platform, entities such as connected vehicles and traffic light controllers could engage in information collecting, data pricing, trading, and decision-making. Simulation results demonstrate that our proposed AI agent-based pricing approach enhances data trading by offering rational prices, as evidenced by the observed improvement in traffic efficiency. This underscores the effectiveness and practical value of DTM, offering new perspectives for the evolution of data markets and smart cities. To the best of our knowledge, this is the first study employing LLMs in data pricing and a pioneering data trading practice in the field of intelligent vehicles and smart cities. 
\end{abstract}

\IEEEpeerreviewmaketitle

\section{Introduction}
With the widespread of Internet-of-Thing (IoT) devices, the amount of collected data is exponentially growing\cite{insights_big_2022}. Data-driven intelligent applications and services have ushered a paradigm shift in various systems, such as public affairs, medical treatments, city management, and traffic \cite{wang_transportation_2023,caoFutureDirectionsIntelligent2022,chen_parallel_2022,li_parallel_2019, lin2023city, duChatChatGPTIntelligent2023, heWKNOCNewDeep2023, liRobustLocalizationIntelligent2023}. The data trading market, as a core productivity in this new era, is predicted to reach USD 655.53 billion by 2029, underscoring its potential to bring value increments to the entire society\cite{insights_big_2022, tripathi_does_2020}. 

\begin{figure}[ht]
\centering
\includegraphics[width=1\linewidth]{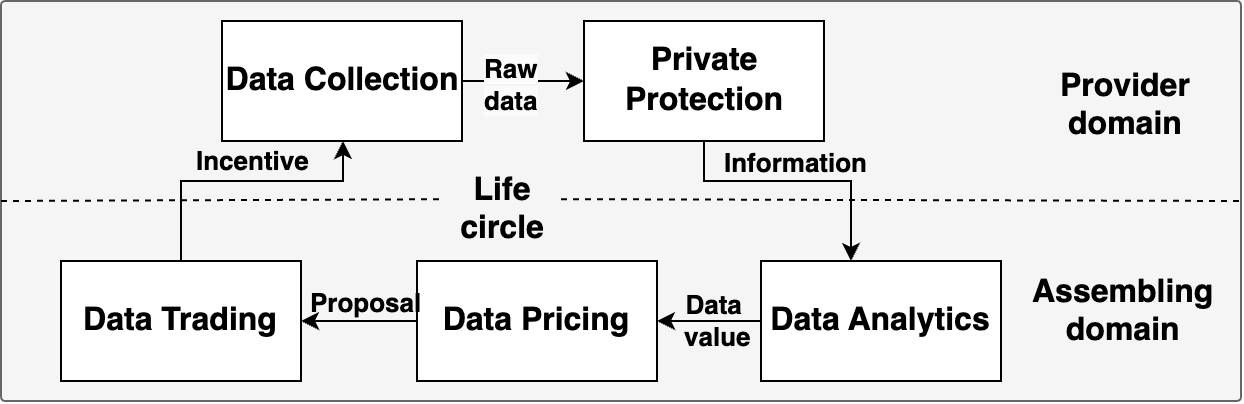}
\caption{Data life cycle in data trading}
\label{life_cycle}
\end{figure}

The data life circle in data trading comprises five essential modules: data collection, private protection, data analytics, data pricing, and data trading, as shown in Fig. \ref{life_cycle}. It is revealed that effective mechanisms of data analytics, data pricing, and data trading in the assembling domain are crucial to stimulate data circulation and unlock the full potential of data productivity, as suggested by Lin et al.\cite{lin_mobility_2023}.     

Existing data pricing models are designed to maximize the provider's profit, fairness, or social welfare \cite{zhang_survey_2020, yuSWDPMSocialWelfareOptimized2023a,yuPursuingEquilibriumMedical2023a, wangDAOMetaControlMetaSystems2022}; whereas tradings are achieved through bilateral agreements, negotiations, or auctions \cite{shen_reinforcement_2020, koutris_query-based_2015, liang2018survey, pei_data_2021, li_novel_2022}. However, these mechanisms are less effective in promoting data circulation in practice \cite{liang_advances_2022}. The reason is twofold: (1) the data value can not be precisely evaluated in complex systems by analytical models; (2) existing trading mechanisms are designed by rational assumptions, whereas the diverse preferences of data buyers and sellers are not considered.      

As one of the most trending data-trading scenarios, the fast development of Intelligent Traffic Systems (ITS) provides an opportunity to address these issues \cite{shen_spatiotemporal_2020, yao_novel_2023, yaoIntegratedTrafficControl2023, yu_identifying_2022,jinCarfollowingModelCalibration2023,zengTrajectoryasaSequenceNovelTravel2023}. Specifically, advanced traffic simulation techniques provide a promising framework for data value evaluation in practice \cite{ wangTransworldngTrafficSimulation2023, wangVerificationValidationIntelligent2022, yangLightingNetIntegratedLearning2023a}. In addition, the rapid development in Large Language Models (LLMs), and Artificial Intelligent (AI) agents powered by LLMs have offered new pathways for simulating tradings in complex systems with the consideration of diverse trading preferences \cite{zhangEmergingTrendsIntelligent2023, wang_chatgpt_2023, fuDeepReinforcementLearning2023,wang_survey_2023}. 

Based on these techniques, we propose a traffic-oriented data trading platform, Data on The Move (DTM), to integrate data analytics, data pricing, and data trading modules in the assembling domain, thus promoting data circulations in practice. The DTM platform incorporates three fundamental modules: trading module, AI agent module, and traffic simulation module. The trading module collects data supply and demand information, distributes trading tasks to the AI agent module, and collects trading results. The AI agent module captures the preferences and the dynamic interactions among various sellers and buyers. The traffic simulation module is applied to evaluate the value of data based on the practical performance of the traffic system. Collaboratively, three modules enable entities such as connected vehicles and traffic light controllers to engage in information collection, data pricing, data trading, and decision-making. In this paper, we focus on the scenario in which connected vehicles sell observed traffic accident data to signal controllers, which utilize this information to optimize control strategies and improve traffic efficiency. 

The key contributions of this paper are twofold: (1) presenting a traffic data trading platform (DTM) that integrates data analytics, pricing, and trading modules by incorporating traffic simulation and AI agent modules; 
(2) proposing an evident-based data value evaluation and pricing mechanism, with an AI agent-based data trading mechanism. The effectiveness of the proposed mechanisms is demonstrated in unlocking data value in traffic systems.

\section{Data on the move platform}\label{system}

Leveraging the common sense capabilities of LLMs and the measurable benefits of traffic systems, we establish the DTM traffic-oriented data trading platform.

\subsection{System structure}

The DTM platform consists of three modules, each dealing with three types of data: traffic system data, trading data, and agent preference data. The platform facilitates interaction between artificial and physical worlds by using real-world data for traffic data trading simulations and leveraging simulation results to guide real-world data trading \cite{li2023parallel}.

The three modules and their functions are:
\begin{enumerate}
    \item \textbf{Traffic Simulation Module}: Simulates traffic flow and generates traffic environmental information.
    \item  \textbf{Data Trading Module}: Collects data demand and supply, distributes trading tasks, and collects trading decisions.
    \item \textbf{Agent Module}: Uses LLMs and the actor model to create connected vehicle and traffic light controller actors; processes data from the traffic simulation and trading modules, and generates trading decisions.
\end{enumerate}

\subsection{Data interaction process}
\begin{figure}[ht]
\vspace{0.1cm}
\centering
\includegraphics[width=0.9\linewidth]{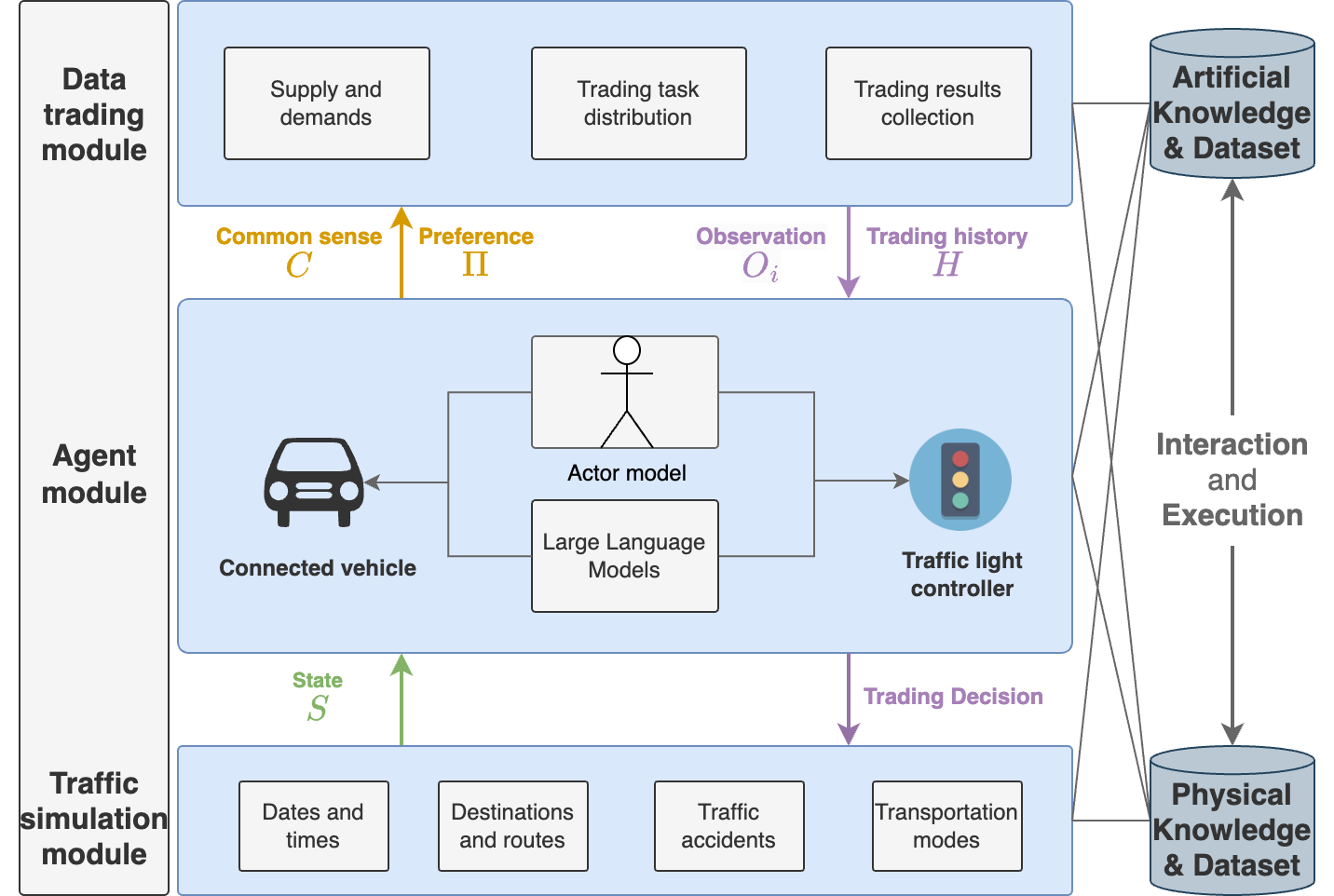}
\caption{Interaction pipeline in the DTM platform}
\label{data_flow}
\end{figure}

The interaction across three modules is shown in Fig. \ref{data_flow}, and involves the following steps: 
\begin{enumerate}
    \item Interaction between the agent module and traffic simulation module: Vehicle agents observe and collect traffic data \(S\) from the traffic simulation module, which serves as the basis for trading.
    \item Interaction between the agent module and the trading module: Vehicle agents provide data to the trading module and trigger data value evaluation; controller agents extract trading data \(H, O\) from the trading module and make decisions based on their preference \(C, \Pi\) and trading data.
    \item Updating traffic and trading modules: The traffic simulation module updates the traffic state based on the trading decision, while the AI agent module updates the trading history data.
\end{enumerate}

\begin{figure}[ht]
\centering
\includegraphics[width=0.9\linewidth]{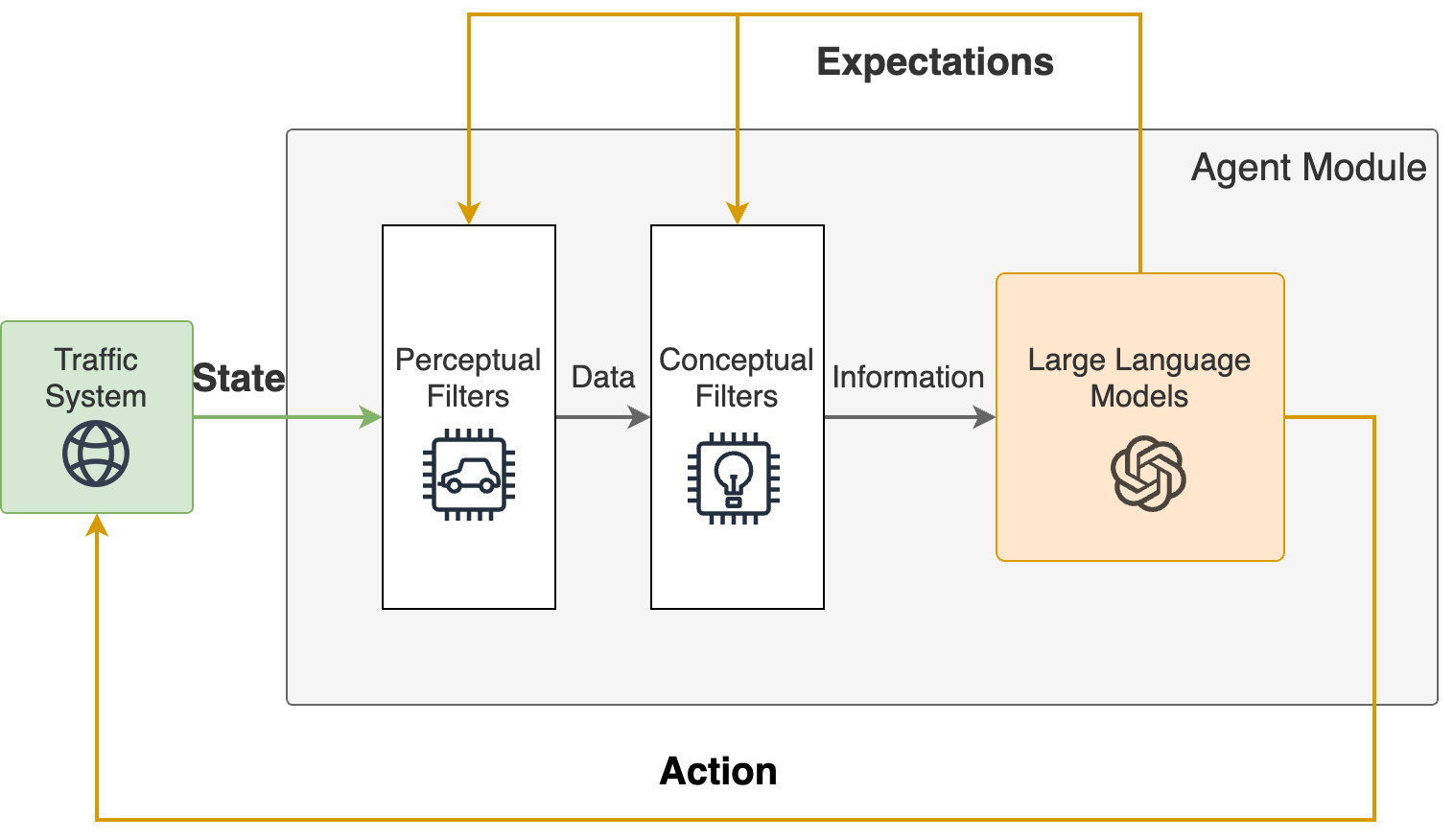}
\caption{Construction and processing flow of the agent module}
\label{agent_module}
\end{figure}

The agent module is the central processor in the DTM platform, extracting states from the traffic simulation and trading modules, processing information, and deriving actions \cite{boisot_data_2004}, as illustrated in Fig. \ref{agent_module}.

\subsection{Evaluation metrics}
To assess the value of data in the traffic system, we measure the change in average waiting time before and after applying the data-driven strategy. The evaluation metric $\Delta\Phi$ is defined as:

\begin{equation}
\Delta\Phi = \Phi_{after} - \Phi_{before}
\end{equation}

where $\Phi$ is the average waiting time:

\begin{equation}
\Phi = \frac{\sum_{i=1}^n \phi_i}{n}
\end{equation}

$\phi_i$: waiting time of vehicle $i$
$n$: total number of observed vehicles

A larger $\Delta\Phi$ indicates a more significant improvement in traffic efficiency, suggesting a higher value of the traded data.

\section{Data pricing trading mechanism} \label{method}

We propose an evident-based pricing and an AI Agent-based data trading mechanism by leveraging the simulation modules and the common sense capability of LLMs in the DTM platform.

\subsection{Value evaluation}
Given the observations from traffic simulation, AI agents can utilize the common sense of LLMs to infer traffic states, analyze the expected improvement in traffic efficiency brought by the data trading, and thereby evaluate the value of traffic data \( V \):

\begin{equation}
    V = \mathbb{E}[\Delta(S)] = llm(C, S)
\label{eq_value}
\end{equation}
where $C$ represents the common sense of LLMs, for example, the understanding of traffic control rules input to LLM via prompts\cite{wang_chatgpt_2023}; $S$ represents the current state of the traffic system, and $llm$ represents the function calling of LLMs.

\subsection{Trading interaction}
After data value evaluation, AI agents participate in several rounds of gaming. Agent $i$ determines its trading strategy $\pi_i$ by considering the data value ($V_i$), personal preferences ($\Pi_i$), observations of other agents' behavior ($O_i$), and trading history ($H_i$):

\begin{equation}
\pi_i = llm(V_i, \Pi_i, O_i, H_i)
\end{equation}

To illustrate the function calling of LLM, examples of prompts are listed in Table \ref{table:prompt_examples}.

\begin{table}[htbp]
\centering
\caption{LLM Prompt and Answer Examples}
\footnotesize 
\begin{tabular}{@{}lp{6.5cm}@{}}
\toprule
Functions & 
\texttt{\{name: "offer\_decision", description: "Decide whether to accept the offer based on the context and available information.", parameters: \{type: "object", properties: \{decision: \{type: "boolean", description: "Acceptance of the offer, where True indicates acceptance and False indicates rejection."\}, reason: \{type: "string", description: "a concise reason why make this decision"\}\}\}, required: ["content"]\}} \\
\midrule
Questions & 
\texttt{\{background: "I am a traffic light controller in an intelligent transportation system, looking to purchase data for controlling support.", risk\_preference: "My risk preference is conservative.", data\_sensitivity: "My data sensitivity is low.", expected\_data\_value: "I expect the data to decrease average delay by 10 seconds", offer\_price: "The data is offered at 12 dollars."\}} \\
\midrule
Responses & 
\texttt{\{content: null, function\_call: \{arguments: "\{decision:false,reason:'The offered data is expected to provide a profit less than the offer price. Considering the conservative risk preference and low data sensitivity, the potential financial benefit does not justify the cost.'\}", name: "offer\_decision"\}, role: "assistant"\}} \\
\bottomrule
\end{tabular}
\label{table:prompt_examples}
\end{table}

\begin{figure}[ht]
\centering
\includegraphics[width=1\linewidth]{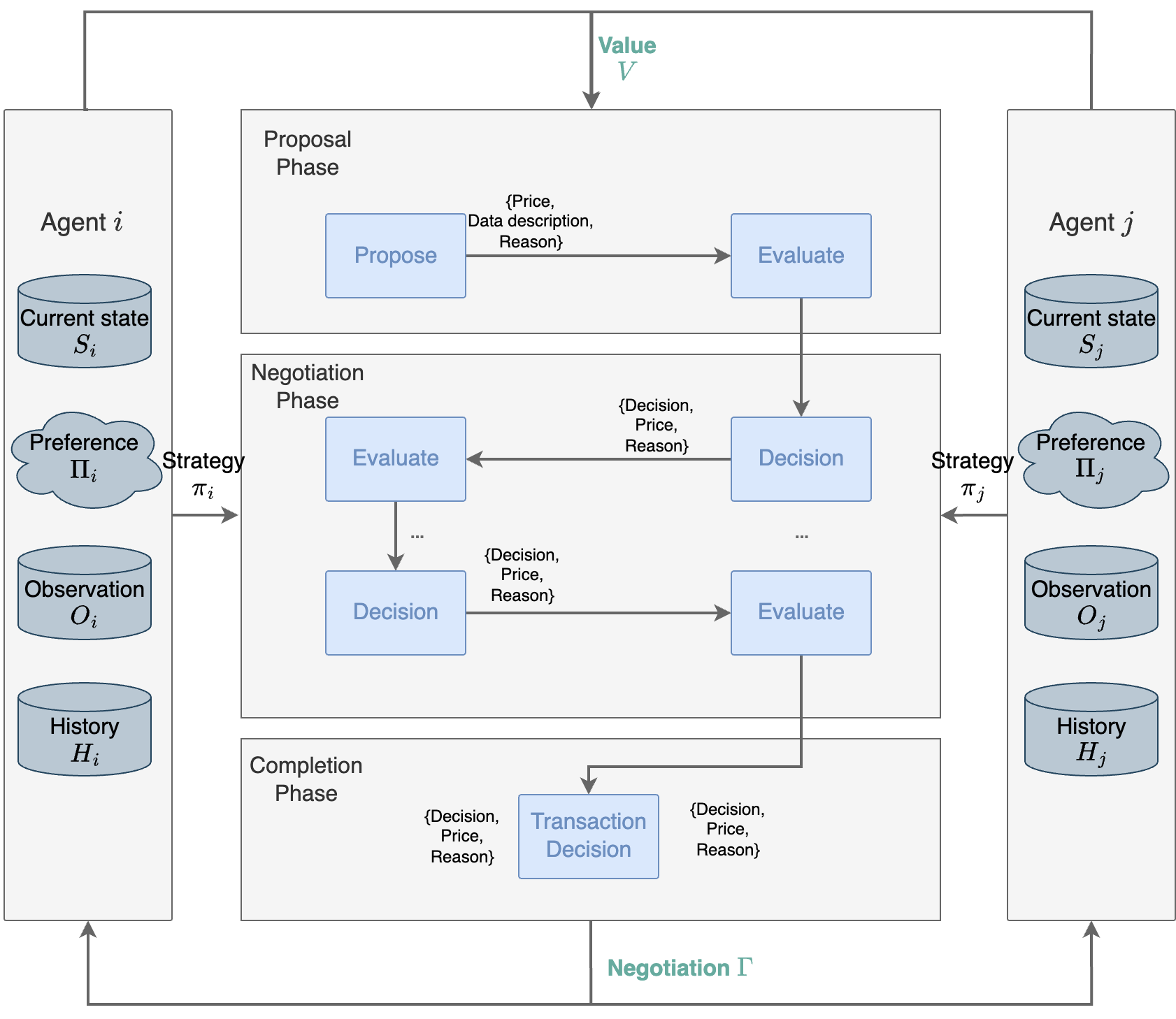}
\caption{AI agent-based data pricing}
\label{AI_agent_pricing}
\end{figure}

Then the agents enter negotiations $\Gamma$:
\begin{equation}
    \Gamma = (N, \{\pi_i\}_{i \in N}, \{U_i\}_{i \in N})
\end{equation}
where $U_i$ represnts the utility of agent $i$, $U_i = U_i(\pi_i, \pi_{-i})$. Agents seek Nash equilibrium through negotiation and strategy adjustment. A schematic of the value assessment and negotiation process is illustrated in Fig. \ref{AI_agent_pricing}.

\subsection{Trading decision}
Combining the results of the value evaluation and multiple round interactions, the final decision and achieved price \( P \) is determined using the pricing function \( f \), reflecting the intrinsic value of the data and the supply-demand relationship in the data market:

\begin{equation}
    P = f(w, V, \Gamma)
\end{equation}
where \( w \) is the weight coefficient, representing the importance of value in the pricing.

\section{Case Study} \label{casestudy}

In this section, we conduct experiments on the DTM platform to evaluate the performance of our proposed AI agent-based data pricing method. 

\subsection{Experiment Settings}
Settings of the DTM platform are introduced as follows. 

\subsubsection{Traffic network and traffic flow}
\begin{figure}[ht]
\centering
\includegraphics[width=1\linewidth]{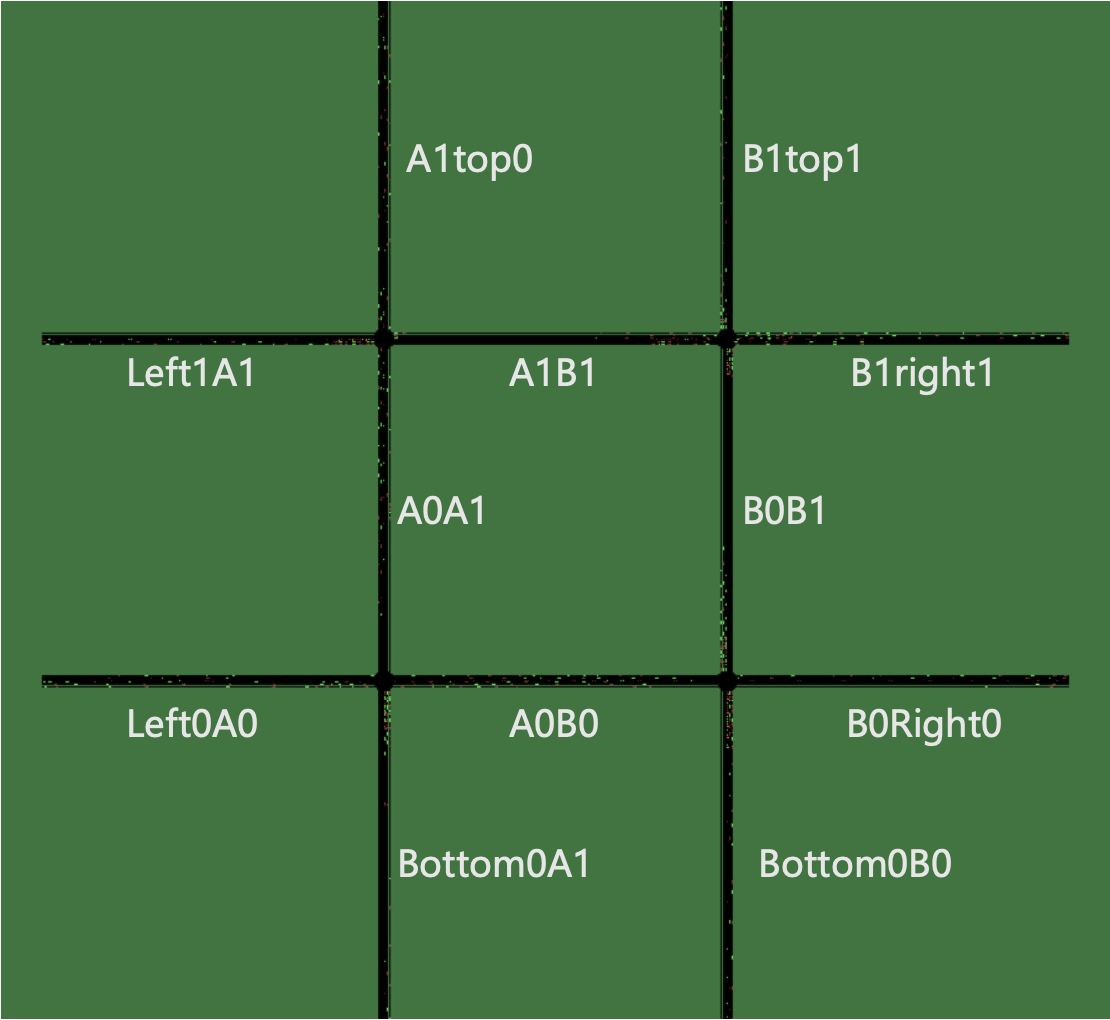}
\caption{Traffic network simulation layout}
\label{traffic_network}
\end{figure}

Network Layout: To simulate typical urban traffic, a grid network with three-lane and 500-meter-long roads is constructed in SUMO (Simulation of Urban MObility), as shown in Fig. \ref{traffic_network}.
Traffic Flow: Different traffic flows (peak and off-peak) are set, with flow ranges from 100 to 500 vehicles/hour, to test the performance of the pricing method under various traffic states.

\begin{figure}[ht]
\centering
\includegraphics[width=1\linewidth]{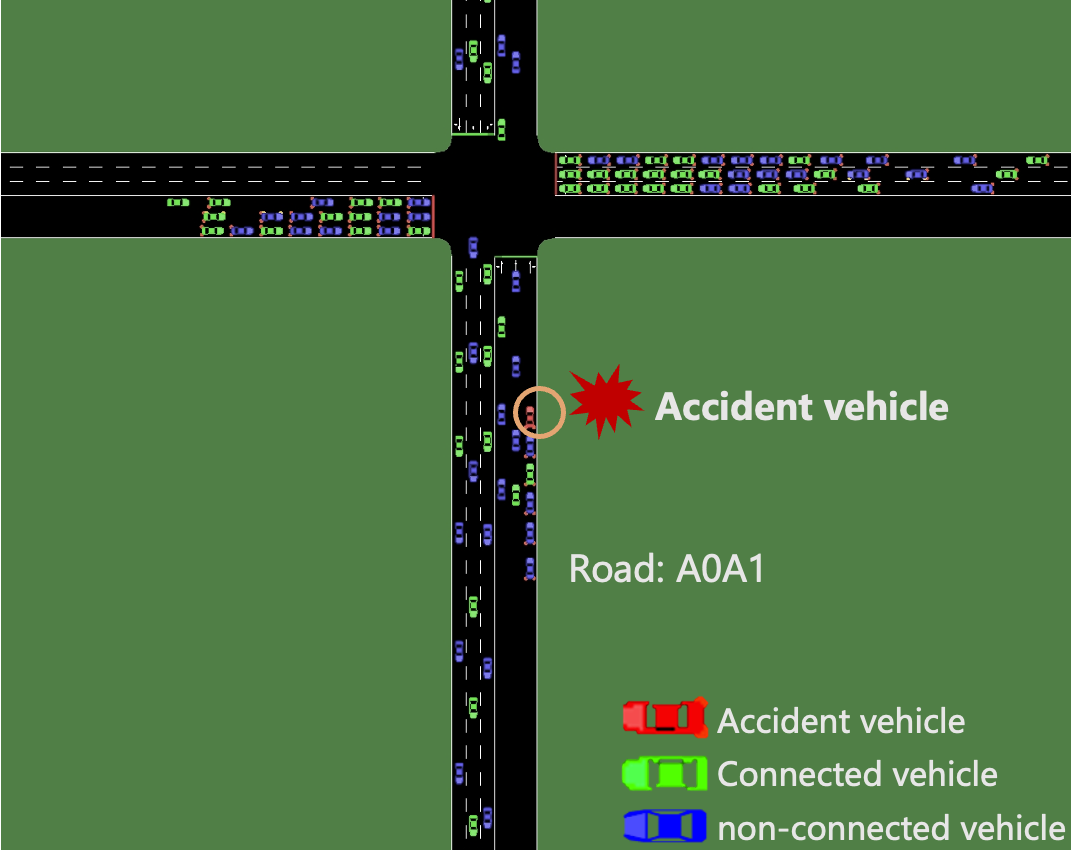}
\caption{Location and impact of traffic accidents}
\label{accident_occurs}
\end{figure}

Accident: Traffic accident data are selected as the trading object. Accidents start from 200 to 700 seconds. The location and impact scope of the accident are shown in Fig. \ref{accident_occurs}.

\subsubsection{Signal Control Strategies}
Baseline Strategy: Traditional static control strategy in the traffic systems.

Data-Driven Strategy: Dynamic control strategy that adjusts traffic light settings (e.g., green light duration and signal cycle) based on the trading data.
   
\subsubsection{Data trading settings}
Connected vehicles (sellers) and traffic light controllers (buyers) participate in data trading, starting with 30 and 100 currency, respectively. Vehicles access the public trading history, collect traffic data every 5 seconds, and propose trades to the controller at trading points, incurring a 1 currency proposal cost. The controller evaluates all proposals and decides whether to accept them. Upon acceptance, the vehicle receives the agreed-upon currency and provides detailed accident data (location, time, severity, and observed traffic flow) to the controller. All agents make decisions using the OpenAI 'gpt-4-1106-preview' model.

\subsubsection{Agent settings}
Preference $\Pi$: Set risk preferences (aggressive, or conservative) and data value sensitivity (high or low). Aggressive agents are more open to risks for potentially higher rewards, while conservative agents prioritize stable outcomes and exercise caution in trading. Agents with high data value sensitivity heavily consider the data's intrinsic value in trading decisions, and vice versa.

With the above settings, a practical traffic data trading platform is built to validate the pricing method and the efficacy of data trading in enhancing traffic efficiency and emergency response.
   
\begin{figure}[ht]
\centering
\includegraphics[width=1\linewidth]{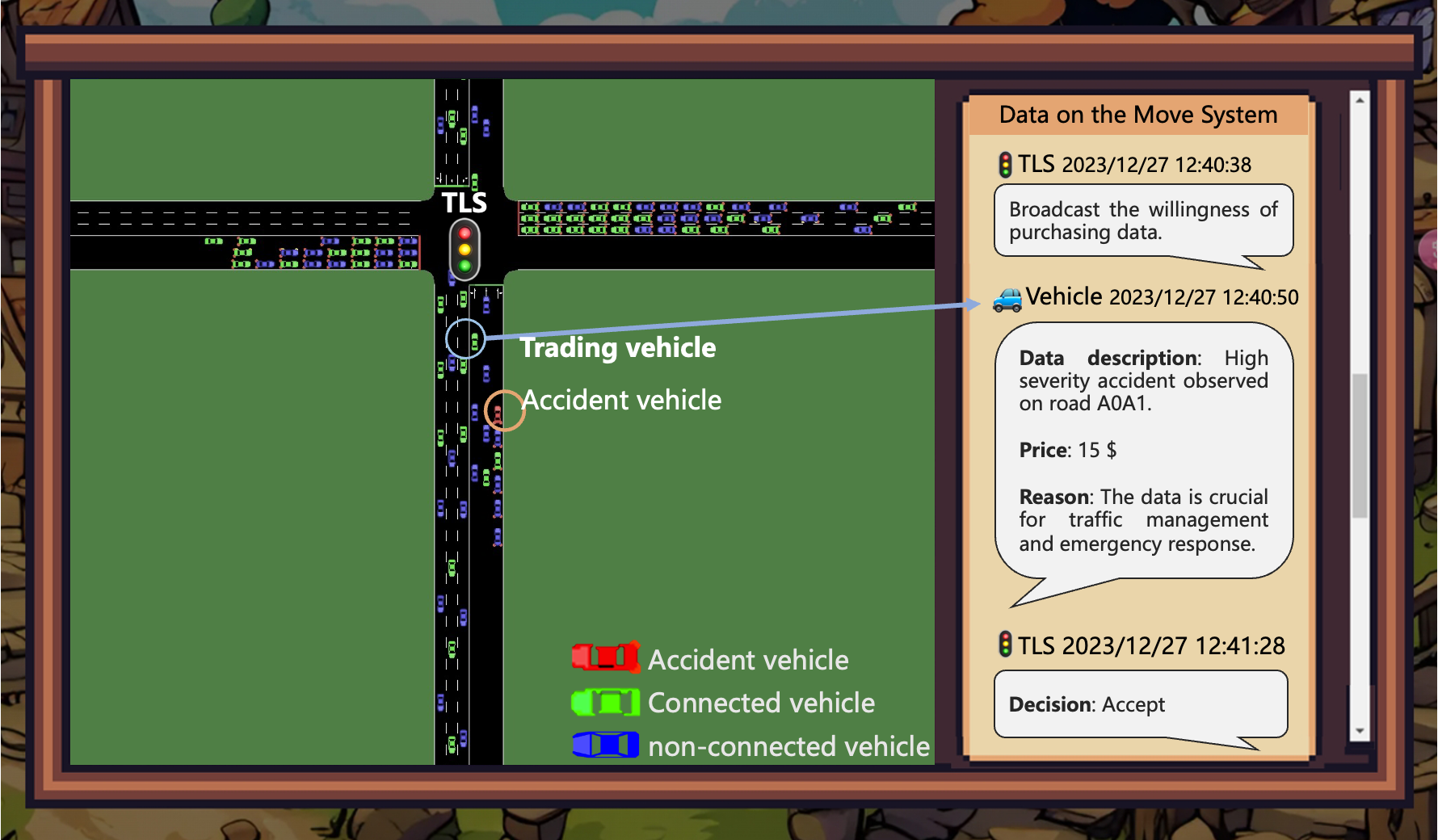}
\caption{Visualization of the DTM platform}
\label{trading_system}
\end{figure}

\subsection{Evaluation Results}
\begin{figure}[ht]
\centering
\includegraphics[width=1\linewidth]{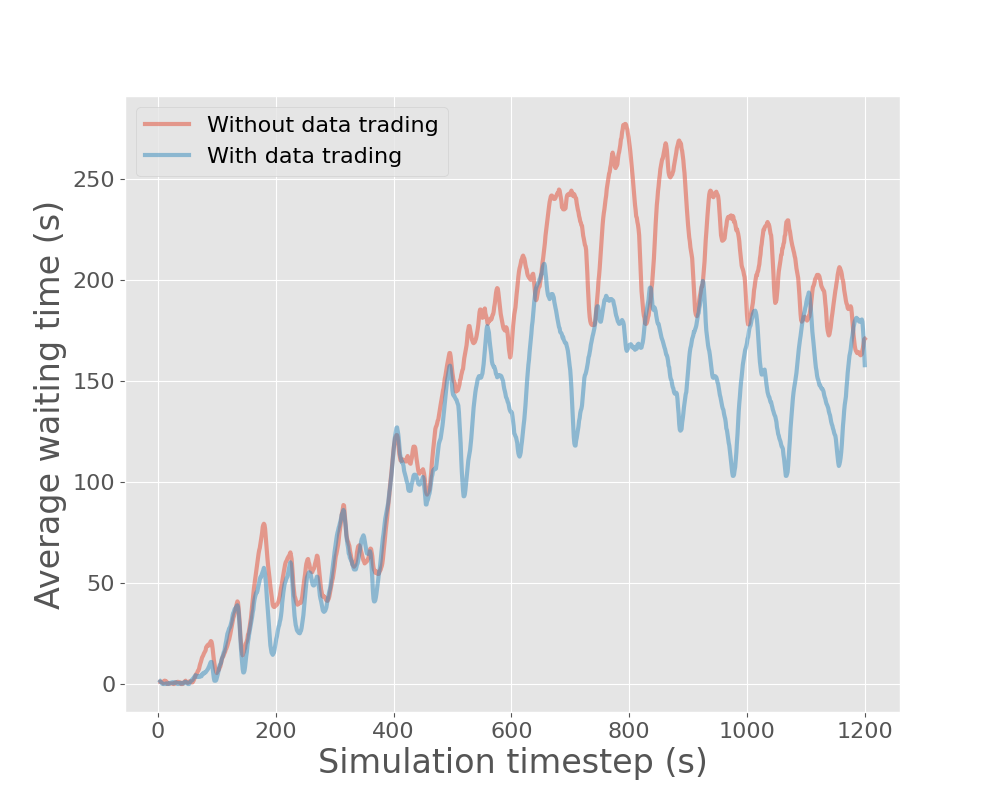}
\caption{Results of average waiting time in the traffic simulation}
\label{average_waiting_time}
\end{figure}

We analyze the data trading under 220 veh/hour traffic flow with conservative and high data sensitivity agents. The controller completes three trades at 230, 235, and 240 seconds using the pricing method. Leveraging the accident data, the controller accurately grasps the location and severity of accidents and adopts the data-driven strategy, prolonging the green light by 3 seconds for the accident entry and shortening another phase by 3 seconds to reduce average waiting time. As shown in Fig. \ref{average_waiting_time}, the controller spends 36 currency on data, achieving a 22.82\% reduction in average waiting time, demonstrating the effectiveness of the data pricing method in enhancing traffic efficiency. 

\begin{figure}[ht]
\centering
\includegraphics[width=1\linewidth]{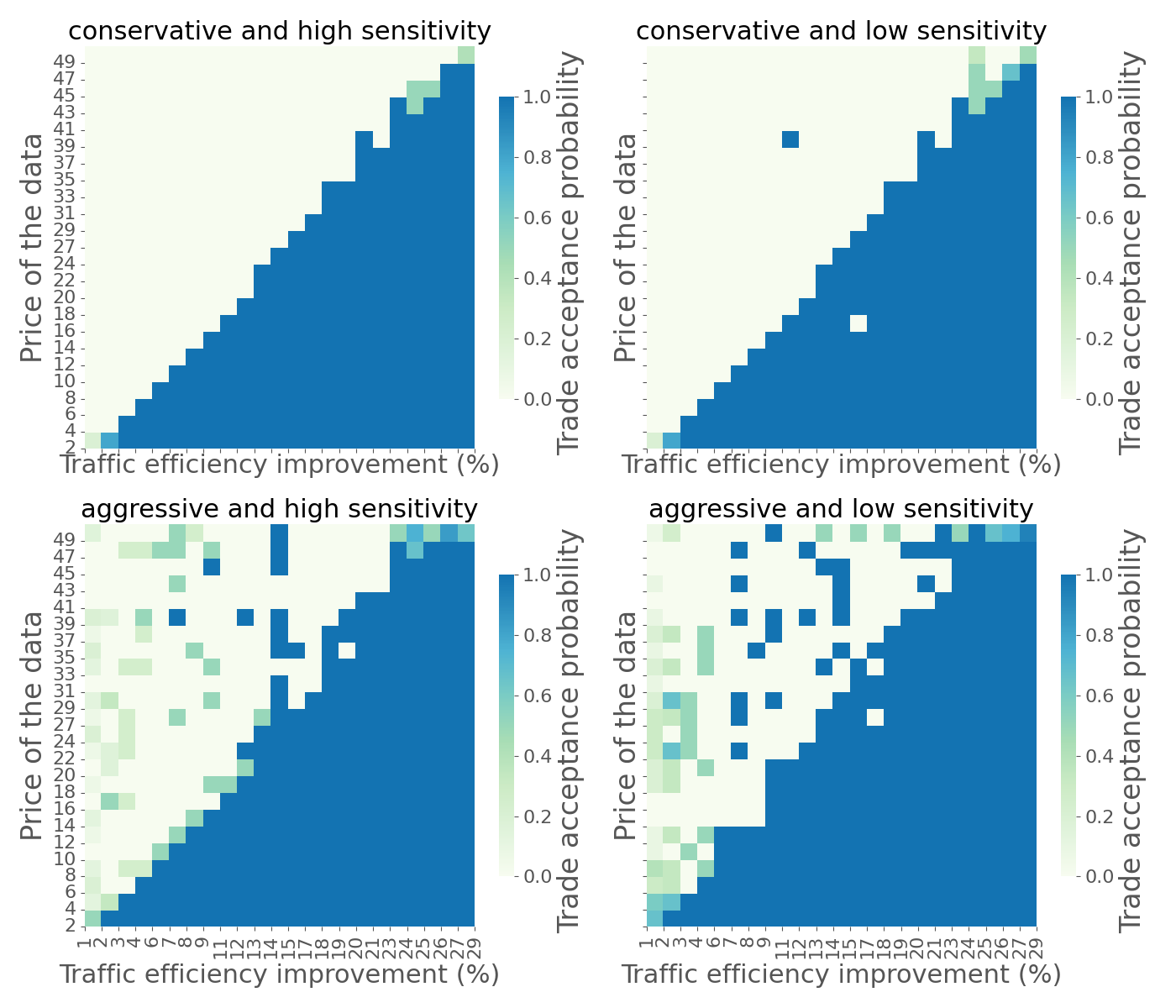}
\caption{Results of trading acceptance probability with different agent preference}
\label{heatmap}
\end{figure}

Furthermore, we evaluated the performance of DTM under various traffic flow and agent preference settings, as shown in Fig. \ref{heatmap}. Results demonstrate that LLMs effectively capture the heterogeneity and potential irrationality in human decision-making. Conservative agents with high data value sensitivity make more rational trading decisions, leading to stable and efficient traffic efficiency improvements at equivalent costs. Conversely, aggressive agents take more risks, accept high-priced data, and have a higher trading success rate. Low-value sensitivity settings decrease LLMs' sensitivity to numerical values, potentially impacting system performance. These findings highlight the importance of considering user preferences when designing and implementing data trading mechanisms in traffic management systems.

\begin{figure}[ht]
\centering
\includegraphics[width=1\linewidth]{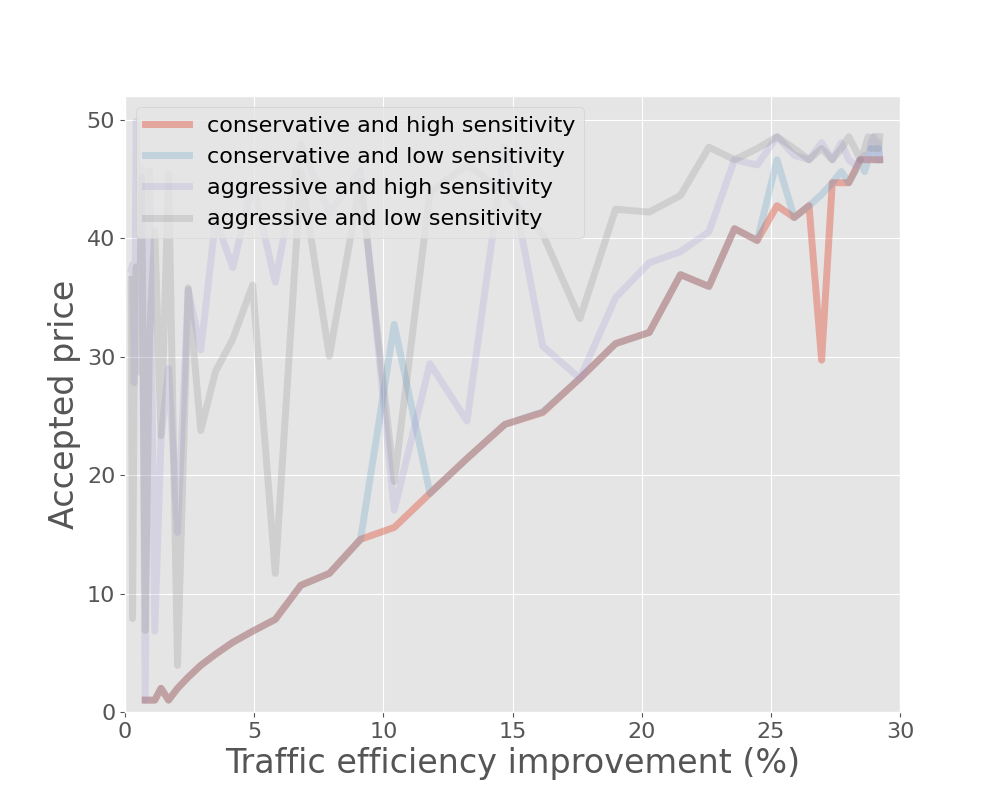}
\caption{Result of pricing vs evaluation metrics with different agent preference}
\label{price_vs_value}
\end{figure}

In Fig. \ref{price_vs_value}, we further compare the data pricing results and the corresponding traffic efficiency improvements under settings, also showing variations in trading decisions among agents. Overall, agents can leverage the common sense advantages of LLMs to precisely understand traffic states, assess the traffic data value, and achieve rational pricing through multi-round interaction. It can be observed that, with a higher performance improvement (up to 29\%), the achieved price with different trading strategies becomes more identical. This finding proved that the DTM platform can facilitate data circulation in practice, which achieves a more diverse trading price according to different strategies when the data value is limited, meanwhile achieving a more regulated price when the data value is high and the performance improvement is fundamental.

The DTM platform proves valuable in supporting pricing validations and effective data pricing through multiple experiments with varied traffic settings and agent preferences. The results highlight the significance of data trading in intelligent transportation systems and provide insights into the development of data markets and smart cities.
\section{Conclusion and future works} \label{conclusion}

In this paper, we propose and validate the DTM platform, offering a trustworthy and practical verification platform for data pricing by combining the common sense capability of LLM with a multi-round game mechanism. The DTM platform consists of traffic simulation, data trading, and Agent modules, using interactions between modules to simulate data trading in traffic systems, providing new perspectives and tools for the research of data markets and smart cities. Moreover, we introduced an AI agent-based data pricing method on DTM, effectively facilitating the circulation of data and unlocking the data value. The results of the simulation experiments confirm that the AI agent-based pricing method can achieve reasonable pricing, demonstrating the realization of effective traffic data circulation via the applicability of DTM.

Considering further improvement and extension, some work awaits future research from the community. For example, advanced technologies such as privacy-preserving, blockchain, and zero-knowledge proofs should be included in the DTM platform to enhance the data privacy protection capabilities of the DTM platform, ensuring the anonymity and immutability of data trading. The successful application of DTM in traffic systems demonstrates its potential in other fields. DTM should be extended to industries such as healthcare and finance, promoting the realization of data value in other fields. Considering future applications, it is necessary to further optimize the computational efficiency to accommodate more complex application scenarios.

\section*{Acknowledgment}
We sincerely appreciate the project managers and engineers at Shanghai Artificial Intelligence Laboratory, including Yiwen Cong, Shanzhe Lei, and Liang Chen, for their valuable contributions to this paper. Their insightful discussions in platform development and project management were essential in the completion of this research.


\end{document}